\documentclass[twocolumn]{article}
\usepackage[T1]{fontenc}
\usepackage{graphics}

\makeatletter

\newcommand{\LyX}{L\kern-.1667em\lower.25em\hbox{Y}\kern-.125emX\@}
\newcommand{\noun}[1]{\textsc{#1}}

\newcommand{\lyxaddress}[1]{
  \par {\raggedright #1 
  \vspace{1.4em}
  \noindent\par}
}

\usepackage{a4}
\input{epsfig.sty}
\def\be{\begin{equation}}
\def\ee{\end{equation}}

\def\epi{\end{picture}}
\def\emi{\end{minipage}}
\def\P{{\rm I\hspace{-0.05cm}P}}

\def\s{{\rm soft }}
\def\soft{{\rm soft}}
\def\semi{{\rm semi}}
\def\R{{\rm I\hspace{-0.05cm}R}}
\def\QCD{{\rm QCD}}
\def\Born{{\rm Born}}

\def\proj{{\rm proj}} 
\def\targ{{\rm targ}}

\makeatother

\begin{document}

\title{\textbf{A Unified Treatment of High Energy Interactions}}

\author{H.J. DRESCHER\protect\( ^{1}\protect \), M. HLADIK\protect\( ^{1,3}\protect \),
S. OSTAPCHENKO\protect\( ^{2,1}\protect \), K. WERNER\protect\( ^{1}\protect \)}

\date{-}

\maketitle

\lyxaddress{\textit{\protect\( ^{1}\protect \) SUBATECH, Universite de Nantes -- IN2P3/CNRS
-- EMN,  Nantes, France }\\
\textit{\protect\( ^{2}\protect \) Moscow State University, Institute of Nuclear
Physics, Moscow, Russia}\\
\textit{\protect\( ^{3}\protect \)now at SAP AG, Berlin, Germany}}

\begin{abstract}
It is well known that high energy interactions as different as electron-positron
annihilation, deep inelastic lepton-nucleon scattering, proton-proton interactions,
and nucleus-nucleus collisions have many features in common. Based upon this
observation, we construct a model for all these interactions, which relies on
the fundamental hypothesis that the behavior of high energy interactions is
universal. 
\end{abstract}

\section{The Universality Hypothesis}

Our ultimate goal is the construction of a model for interactions of two nuclei
in the energy range between several tens of GeV up to several TeV per nucleon
 in the center-of-mass system. Such nuclear collisions are very complex, being
 composed of many components, and therefore some strategy is needed to construct
a reliable model. The central point of our approach is the hypothesis, that
the behavior of high energy interactions is universal (universality hypothesis).
So, for example, the hadronization of partons in nuclear interactions follows
the  same rules as the one in electron-positron annihilation; the radiation
of  off-shell partons in nuclear collisions is based on the same principles
as the  one in deep inelastic scattering. 

The structure of nucleus-nucleus scattering is expected to be as follows: there
are elementary interactions between individual  nucleons, realized via parton
ladders, where the same nucleon may  participate in several of these elementary
interactions. Although such  diagrams can be calculated in the framework of
perturbative QCD, there  are quite a few problems : important cut--offs have
to be chosen, one has to  choose the appropriate evolution variables, one may
question the validity of the  ``leading logarithmic approximation'', the coupling
of the parton ladder to the nucleon is  not known, the hadronization procedure
is not calculable from first principles  and so on. So there are still many
unknowns, and a more detailed study is  needed.  

Our starting point is the universality-hypothesis, saying that  \textit{the
behavior of high-energy interactions is universal} \cite{wer97}. In this case
all the details of nuclear interactions can be determined by studying  simple
systems in connection with using a modular structure for modeling  nuclear scattering.
One might think of proton-proton scattering representing a   simple system,
but this is already quite complicated considering the fact  that  we have in
general already several ``elementary interactions''.

\section{The semihard Pomeron}

\begin{figure}
{\par\centering \resizebox*{0.15\textwidth}{!}{\includegraphics{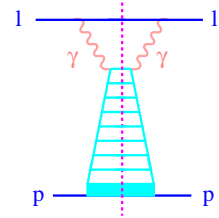}} \par}

\caption{The  diagram representing deep inelastic  lepton-proton scattering. \label{dis}}
\end{figure}

Let us call an elementary interaction in proton-proton scattering at high energies
``semihard Pomeron''. In order to investigate the structure of the semihard
Pomeron, we turn to an even simpler  system, namely deep inelastic lepton-nucleon
scattering (DIS).   In figure \ref{dis} we show the cut diagram representing
lepton-proton  scattering: a photon is exchanged between the lepton and a quark
of the  proton, where this quark represents the last one in a ``cascade''
of partons  emitted from the nucleon. So the hadronic part of the diagram is
essentially a parton ladder. 
In the leading logarithmic approximation (LLA) the virtualities of the partons
are  ordered such that the largest one is close to the photon \cite{rey81,alt82}.

\begin{figure}
{\par\centering \resizebox*{0.3\textwidth}{!}{\includegraphics{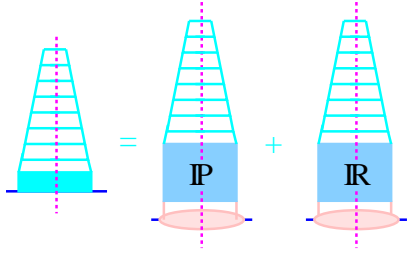}} \par}

\caption{The total contribution, being  the sum of IP-contribution and IR-contribution.\label{ftot}}
\end{figure}

\begin{figure}
{\par\centering \resizebox*{0.9\columnwidth}{!}{\includegraphics{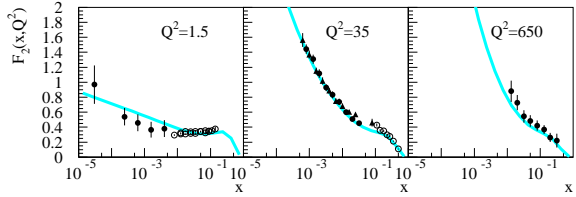}} \par}

\caption{The structure function \protect\( F_{2}\protect \)for different values of
\protect\( Q^{2}\protect \) together with experimental data from H1 \cite{h1-96a},
ZEUS \cite{zeus96}and NMC \cite{nmc95}.\label{ep1}}
\end{figure}

\begin{figure}
{\par\centering \resizebox*{0.2\textwidth}{!}{\includegraphics{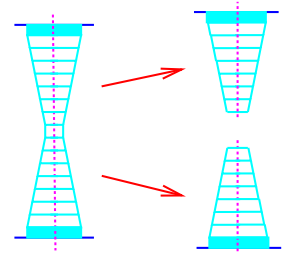}} \par}

\caption{The universality hypothesis implies that the upper and the lower part of the
Pomeron diagram are identical to the hadronic part of the diagram for DIS. \label{cpomdis}}
\end{figure}

Let us first investigate the so-called structure function \( F_{2} \),  representing
the hadronic part of the DIS cross section, i.e. the diagram of fig. \ref{dis}
but without the lepton and photon lines. In DGLAP approximation, we may write
\( F_{2} \) as
\begin{equation}
F_{2}(x,Q^{2})=\sum _{j}e_{j}^{2}\, x\, f^{j}(x,Q^{2})
\end{equation}
 with 
\[
f^{j}=\sum _{i}\varphi ^{i}\otimes E^{ij}_{\mathrm{QCD}}.\]
Here, \( E^{ij}_{\mathrm{QCD}}(x,Q^{2}_{0},Q^{2}) \) is the QCD evolution function,
representing the evolution of a parton cascade from scale \( Q^{2}_{0} \) to
\( Q^{2} \), being calculated based on the DGLAP evolution equation. We solve
this equation numericly in an iterative fashion. The function \( \varphi ^{i} \)
is the initial parton distribution (at scale \( Q^{2}_{0} \)), assumed to be
of the form
\[
\varphi ^{i}=C_{\P }\otimes E^{i}_{\s \P }+C_{\R }\otimes E^{i}_{\s \R }.\]
So we have two contributions, represented by a soft Pomeron and a Reggeon respectively.
In any case we adopt a form \( C\otimes E_{\mathrm{soft}} \), where \( E_{\mathrm{soft}} \)
represents the soft Pomeron/Reggeon, and where \( C \) is the coupling between
the Pomeron/Reggeon and the nucleon. So the parton ladder is coupled to the
nucleon via a soft Pomeron or Reggeon, see fig. \ref{ftot}. Here, we regard
the soft Pomeron (Reggeon) as an effective description of a parton cascade in
the region, where perturbative methods are inapplicable. A similar construction
was proposed in \cite{tan94, lan94}, where a t-channel iteration of soft and
perturbative Pomerons was considered. We call \( E_{\s } \) also the soft evolution,
to indicate that we consider this as simply  a continuation of the QCD evolution,
however, in a region where perturbative  techniques do not apply any more.  
 Some results of our calculations for  \( F_{2}(x,Q^{2}) \) are shown in fig.
\ref{ep1} together with experimental data from H1 \cite{h1-96a}, ZEUS \cite{zeus96}
and NMC \cite{nmc95}.
 We are now in a position to write down the expression \( G_{\semi } \) for
a \textit{cut semihard Pomeron}, representing an elementary inelastic interaction
in \( pp \) scattering. We can divide the corresponding  diagram  into three
parts: we have the process involving the highest parton virtuality in the middle,
and  the upper and lower part representing each an ordered parton ladder coupled
to  the nucleon. According to the universality hypothesis, the two latter parts
are  known from studying deep inelastic scattering, representing each the hadronic
 part of the DIS diagram, as shown in fig. \ref{cpomdis}.
 For given impact parameter \( b \) and given energy squared \( s \), the
complete diagram is therefore given as 
\begin{equation}
\int dx^{+}dx^{-}\, G_{\semi }(x^{+},x^{-}),
\end{equation}
 with

\begin{eqnarray}
 &  & G_{\semi }(x^{+},x^{-})=\nonumber \\
 &  & \, \, \, \sum _{IJ}\, C_{I}(x^{+})\, C_{J}(x^{-})\, \nonumber \\
 &  & \, \, \, \sum _{ijkl}\, \int du^{+}du^{-}dQ^{2}\, [E^{k}_{\soft \, I}\otimes E^{ki}_{\QCD }](u^{+})\\
 &  & \, \, \, [E^{l}_{\soft \, J}\otimes E^{lj}_{\QCD }](u^{-})\, {d\sigma ^{ij}_{\Born }\over dQ^{2}}(u^{+}u^{-}x^{+}x^{-}s,Q^{2}).\nonumber \label{gsemi} 
\end{eqnarray}
 The variables \( I \) and \( J \) may take the values \( {\textrm{I}\! \textrm{P}} \)
and \( {\textrm{I}\! \textrm{R}} \). This is the expression corresponding to
a \( semihard \) \( Pomeron \). 

In addition to the semihard Pomeron, one has to consider the expression representing
the purely soft contribution \( G_{\soft } \) \cite{hla98}. The complete contribution
is then the sum \( G=G_{\semi }+G_{\soft } \).

\section{Multiple Scattering in Nucleus-Nucleus collisions}

\begin{figure}
{\par\centering \resizebox*{1\columnwidth}{!}{\includegraphics{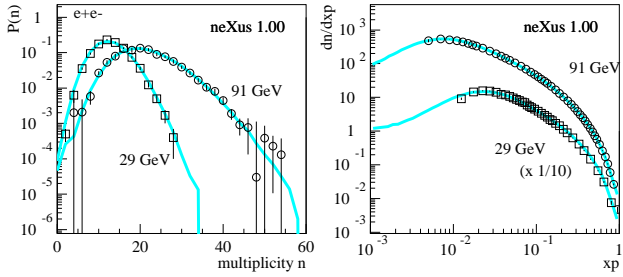}} \par}

\caption{Multiplicity (left)  and \protect\( x_{p}\protect \) distributions (right)
at 29 GeV (lower curves) and 91 GeV (upper curves). The 29 GeV results have
been divided by ten, to separate the two curves.\label{expl-ee}}
\end{figure}

We define a consistent multiple scattering theory for nucleus-nucleus scattering
(including proton-proton) as follows:

\begin{itemize}
\item Any pair of nucleons may interact via the exchange of any number of cut or uncut
Pomerons of any kind (Reggeons, soft Pomerons, semihard Pomerons).
\item The total cross section is the sum of all such ``Pomeron configurations''.
\end{itemize}
As usual, cut Pomerons represent contributions of partial nucleon-nucleon interactions
into real particle production, whereas uncut ones are the corresponding virtual
processes (screening corrections) \cite{agk73, wer93}. With each cut Pomeron
 contributing a factor \( G \), each uncut one a factor \( (-G) \)(the Pomeron
amplitude is assumed to be imaginary), and each remnant contributing a factor
\( F_{\mathrm{proj}} \) or \( F_{\mathrm{targ}} \), one gets
\begin{eqnarray}
\sigma _{\mathrm{inel}} & = & \int dT_{AB}\sum _{m_{1}l_{1}}\ldots \sum _{m_{AB}l_{AB}}\, \prod _{k=1}^{AB}\left\{ \frac{1}{m_{k}!}\frac{1}{l_{k}!}\right\} \int \, dX\nonumber \\
 &  & \prod _{k=1}^{AB}\left\{ \prod _{\mu =1}^{m_{k}}G(x_{k,\mu }^{+},x_{k,\mu }^{-})\, \prod _{\lambda =1}^{l_{k}}-G(\tilde{x}_{k,\lambda }^{+},\tilde{x}_{k,\lambda }^{-})\right\} \nonumber \\
 &  & \prod _{i=1}^{A}F_{\proj }\left( x^{\mathrm{R}+}_{i}\right) \, \prod _{j=1}^{B}F_{\targ }\left( x^{\mathrm{R}-}_{j}\right) \label{s1} 
\end{eqnarray}
 where \( \int dT_{AB} \) represents the integration over impact parameters
of projectile and target nucleons with the appropriate weight given by the so-called
thickness functions, \( \int dX \) represents the integration over all momentum
fraction variables, and \( x_{i}^{\mathrm{R}+} \)and \( x^{\mathrm{R}-}_{j} \)represent
remnant momentum fractions. Equation (\ref{s1}) should be considered symbolic,
in reality, there  appear also transverse momentum variables and one has to
take into account the fact that the case of zero cut Pomerons is somewhat complicated:
one has elastic and diffractive interactions as well as cuts  between nucleons.
All this is taken into account in the numerical calculations \cite{hla98}.
Eq. (\ref{s1}) is the basic formula of our approach, it serves not only to
generate Pomeron configurations (how many Pomerons of which type are exchanged),
it serves also as basis to generate partons. The final step amounts to transform
partons into hadrons (hadronization). This is done according to the so-called
kinky string method, which has been tested very extensively  for electron-positron
annihilation. As an example, we present in fig. \ref{expl-ee}  multiplicity
and \( x_{p} \) distributions at 29 and 91 GeV together with data from \cite{der86}
and \cite{ale96}, where \( x_{p} \) is the momentum fraction of a particle.
In a similar fashion, we reproduce data concerning inclusive spectra for  individual
hadrons. 

\section{Results}

Let us now discuss some results concerning particle production, first in  deep
inelastic lepton-nucleon scattering. 
\begin{figure}
{\par\centering \resizebox*{1\columnwidth}{!}{\includegraphics{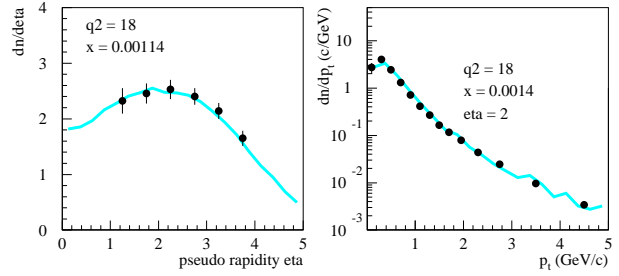}} \par}

\caption{Pseudo-rapidity distributions in DIS for charged particles (left) or for charged
particles with \protect\( p_{t}>\protect \) 1 GeV (right). Data are from \cite{h1-96b}.\label{ep2x1}}
\end{figure}
 In fig. \ref{ep2x1}, we present pseudo-rapidity distributions and transverse
momentum spectra for  \( 1.5<\eta <2.5 \). We compare our simulations (lines)
with H1 data (points). We applied the same acceptance cuts as done in the experiment. 

\begin{figure}
{\par\centering \resizebox*{0.9\columnwidth}{!}{\includegraphics{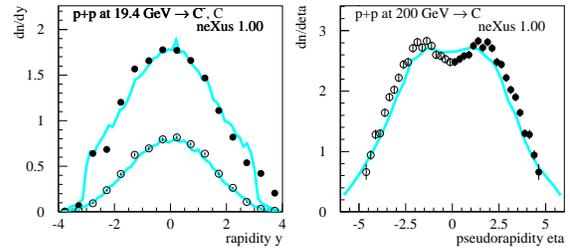}} \par}

\caption{Rapidity or pseudorapidity distributions at 19.4 and 200 GeV. The data are
from \cite{dem82, UA5-86}\label{spectra-1}}
\end{figure}

We are now going to discuss a few results for \( pp \) scattering. In fig.
\ref{spectra-1}, 
we present on the left-hand-side rapidity spectra of  charged particles (upper
curve) and negatively charged particles (lower curve) at a center-of-mass energy
of 19.4 GeV. On the right-hand-side, we show pseudorapidity  spectra of charged
particles at 200 GeV. 
\begin{figure}
{\par\centering \resizebox*{1\columnwidth}{!}{\includegraphics{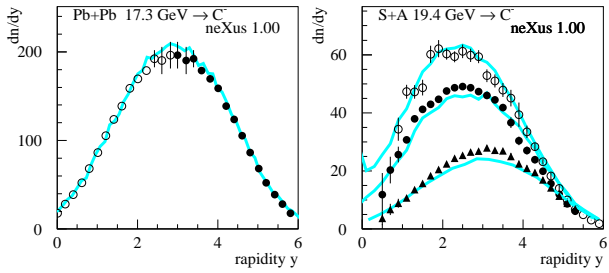}} \par}

\caption{Rapidity distribution of negatives for Pb+Pb (left) and S+S, S+Ag, S+Au (right).
The data are from \cite{NA49-98, NA35-94}\label{aa2}}
\end{figure}

In order to treat nucleus-nucleus collisions, one needs to include secondary
interactions. This cannot be done within the theoretical framework discussed
for far. So we proceed in two steps:

\begin{itemize}
\item We first treat the ``primary interactions'', according to the procedures discussed
above. This is a consistent multiple scattering approach, fully compatible with
proton-proton scattering and deep inelastic lepton-nucleon scattering.
\item We then reconsider the event, to perform the ``secondary interactions'', i.e.
to check whether at least three particles are close to each other to form droplets,
or if two particles are close to each other to perform a hadron-hadron interaction. 
\end{itemize}
It goes beyond the scope of this paper to discuss the details of ``secondary
scattering'', this will be left to a future publication, where we also are going
to present a detailed comparison with data. In fig. \ref{aa2}, we show rapidity
distributions of negatives for Pb+Pb and for S+S, S+Ag, S+Au at SPS. In general,
we obtain an agreement with the data on the level of 5 - 10 \%.

\section{Summary}

Based on the universality hypothesis, we constructed a theoretically consistent
approach to high energy interactions as different as deep inelastic lepton-nucleon
scattering, proton-proton scattering and nucleus-nucleus scattering. Both, interactions
of nucleons or nuclei, are complex in the sense that the cut Feynman diagrams
contributing to the total cross section are composed of ``subdiagrams'' representing
elementary interactions between nucleons. These subdiagrams are called semihard
Pomerons, they are parton ladders coupled to the nucleons. Each subdiagram can
be divided into two parts, each one representing a diagram, which can be studied
via deep inelastic lepton-nucleon scattering. This is very important, because
in this way we can study the soft coupling of the parton ladder to the nucleon,
not being known from QCD calculations, fit the parameters of parameterizations
by comparing to lepton scattering data, and have in this way no freedom any
more in proton-proton or nucleus-nucleus scattering.

This work has been funded in part by the IN2P3/CNRS (PICS 580) and the Russian
Foundation of Fundamental Research (RFFI-98-02-22024).

\end{document}